\documentclass[%
preprint,
amsmath,amssymb,
aps,
]{revtex4-2}

\usepackage{graphicx}% Include figure files
\usepackage{dcolumn}% Align table columns on decimal point
\usepackage{bm}% bold math
\usepackage[dvipsnames]{xcolor}

\definecolor{a}{RGB}{170, 80, 133}
\definecolor{b}{RGB}{185, 110, 85}
\definecolor{c}{RGB}{159, 159, 63}
\definecolor{d}{RGB}{129, 77, 174}
\definecolor{e}{RGB}{95, 95, 191}
\definecolor{f}{RGB}{77, 129, 174}
\definecolor{g}{RGB}{110, 185, 85}
\definecolor{h}{RGB}{80, 170, 133}

\definecolor{a}{RGB}{255, 255, 255}
\definecolor{b}{RGB}{255, 255, 255}
\definecolor{c}{RGB}{255, 255, 255}
\definecolor{d}{RGB}{255, 255, 255}
\definecolor{e}{RGB}{255, 255, 255}
\definecolor{f}{RGB}{255, 255, 255}
\definecolor{g}{RGB}{255, 255, 255}
\definecolor{h}{RGB}{255, 255, 255}

\newcommand{\Rey}{\mathrm{Re}}
\newcommand{\We}{\mathrm{We}}
\newcommand{\Ca}{\mathrm{Ca}}

\newcommand{\Emax}{\mathcal{E}_\text{max}}
\newcommand{\Einf}{\mathcal{E}_\infty}
\newcommand{\Amax}{A_\text{max}}
\newcommand{\Vmax}{\mathcal{V}_\text{max}}
\newcommand{\Estat}{\mathcal{E}_\text{stat}}
\newcommand{\Edyn}{\mathcal{E}_\text{dyn}}

\newcommand{\etal}{\textit{et al.}~}
\newcommand{\eq}[1]{Eq. \eqref{#1}}
\newcommand{\fig}[1]{Fig. \ref{#1}}

\begin{document}

	\title{Hydrodynamic constraints on the energy efficiency of droplet electricity generators}% Force line breaks with \\
	%	\thanks{A footnote to the article title}%
	
	\author{Cui Wang}

	\author{Jia Zhou}

	\author{Antoine Riaud}
	\affiliation{State Key Laboratory of ASIC and System, School of Microelectronics, Fudan University, Shanghai 200433, China}
	\homepage{http://homepage.fudan.edu.cn/ariaud/}
	\email{antoine\_riaud@fudan.edu.cn}

	\author{Wanghuai Xu}

	\author{Zuankai Wang}
	\affiliation{Department of Mechanical Engineering, City University of Hong Kong, Hong Kong 999077, China}
	\email{zuanwang@cityu.edu.hk}
	
	\date{\today}% It is always \today, today,
	
	\begin{abstract}
		Electric energy generation from falling droplets has seen a hundred-fold rise in efficiency over the past year. However, even these newest devices can only extract a small portion of the droplet energy. In this paper, we theoretically investigate the contributions of hydrodynamic and electric losses in limiting the efficiency of droplet electricity generators (DEG). Noting that the electro-mechanical energy conversion occurs during the recoil that immediately follows droplet impact, we identify three limits on existing droplet electric generators: (i) the impingement velocity is limited in order to maintain the droplet integrity; (ii) much of droplet mechanical energy is squandered in overcoming viscous shear force with the substrate; (iii) insufficient electrical charge of the substrate. Of all these effects, we found that up to 83\% of the total energy available was lost by viscous dissipation during spreading.  Minimizing this loss by using cascaded DEG devices to reduce the droplet kinetic energy may increase future devices efficiency beyond 10\%.
	\end{abstract}
	
	\maketitle
	
	%\tableofcontents
	
	\section{Introduction}

	%Despite an impressive 100-fold increase within the past few years, the efficiency of droplet electric generators ($\simeq10$ \%) still lags far behind dams ($\simeq100$ \%). 
	
	%difference of scale = turbulent vs laminar, laminar dissipates more (for instance drag force on a sphere)
	
	Droplet electricity generators (DEG) are designed to harvest the kinetic energy of rain droplets to power small wireless sensors. Despite a 100-fold increase in efficiency over the past few years \cite{xu2020a,wu2020energy,xu2020fusion}, even state-of-the-art devices only recover 10\% of the kinetic energy of water \cite{wu2020charge}, as opposed to the nearly 100\% efficiency achieved by hydroelectric dams.
	
\begin{figure}
		\includegraphics[width=80 mm]{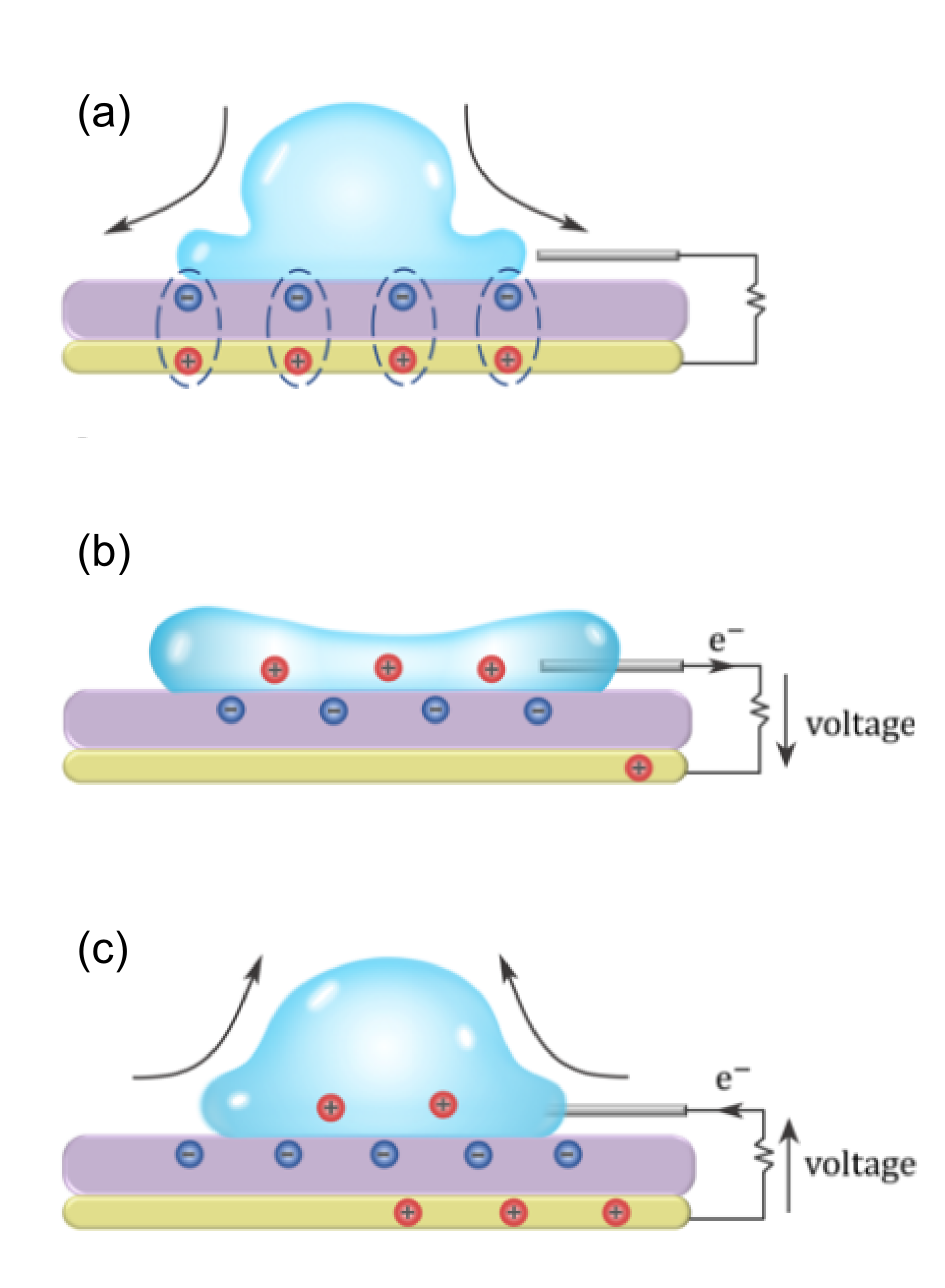}
		\caption{Droplet electric generator with charge circulation \cite{xu2020a,wu2020charge,wu2020electrically}. \textbf{(a)} The substrate is initially charged by the impingement of many droplets or other electrical forcing \cite{wu2020electrically} and forms charge pairs on each side of the substrate. \textbf{(b)} Upon contact, the substrate capacitor is discharged through the load and the liquid. \textbf{(c)} During the recoil, trapped charges in the polymer are left behind, so that positive charges move back to the ITO to restore the charge pairs. \label{fig: schematic}}
	\end{figure}

	Unlike dams, which extract energy from the mechanical work of water on the hydro-turbines, DEG, and more broadly triboelectric nanogenerators (TENG) harvest energy from charges accumulated on surfaces which are then used to drive an electric current through an external circuit by electrostatic induction \cite{lin2014harvesting}. In the case of DEG, the charges are spontaneously created by water at the surface of polymers \cite{yatsuzuka1994electrification,zimmermann2010hydroxide,banpurkar2017spontaneous,stetten2019slide} by contact electrification \cite{lin2013water}, a process which can be intensified by applying a voltage across the polymer layer \cite{wu2020electrically}. In the latest studies, a grounded metallic electrode is placed underneath the polymer and is connected to a small metallic strip on the top (see \fig{fig: schematic}). According to the present understanding, \cite{wu2020energy,wang2020dynamics}, this sandwiched structure then behaves as a biased capacitor. Upon contact with water, the capacitor is discharged through the load, which releases the electrostatic energy that was stored previously \cite{wu2020energy}. Meanwhile, mobile charges accumulate at the water-polymer interface. When the droplet recedes, those charges are detached from the interface and forced to return to the bottom electrode \cite{wu2020charge}.  While this picture predicts the transfer of charges through the DEG with a remarkable accuracy \cite{wu2020energy}, it does not consider the hydrodynamic side of the picture. Yet, the harvested electrical energy accounts at best for 10\% of the initial droplet energy, meaning that, in our present understanding, at least 90\% of the droplet energy is unaccounted for.

	A comprehensive DEG model would consider (i) the electrical process at play, (ii) the hydrodynamics (iii) the electrochemical charge stability and (iv) the electrohydrodynamic coupling. Since only 10\% of the DEG energy is electrical, we neglect the electrohydrodynamic coupling and focus mainly on the hydrodynamic process. In this simplified view, the DEG hydrodynamics are exactly those of a droplet impacting an inclined plane. Even in this elementary picture, three effects compete to dominate the droplet dynamics: inertia, surface forces, and viscous dissipation. For a spherical droplet of radius $a$, density $\rho$, viscosity $\mu$, surface tension $\gamma$ falling at a velocity $U_0$, the ratio of kinetic energy to viscous work is approximately the Reynolds number $\Rey = \frac{\rho a U_0}{\mu}$, while the ratio of kinetic energy to surface energy is connected to the Weber number $\We = \frac{\rho {U_0}^2 a}{\gamma}$. During impacts, the liquid spreads into a thin lamella on the solid where most of the energy conversion occurs \cite{wildeman2016spreading}. It was noticed quite early that a large fraction of the droplet energy is lost during impact  \cite{chandra1991collision}. This loss was attributed to viscous dissipation within the lamella \cite{chandra1991collision,pasandideh1996capillary,ukiwe2005on} until experiments of low-viscosity drop impacts on super-hydrophobic and slippery substrates suggested that a large fraction of the energy was actually converted to internal kinetic energy (akin to turbulence) \cite{richard2002contact,clanet2004maximal,biance2006elasticity}. Several recent works have since attempted to bridge the gap between these two regimes \cite{laan2014maximum,roisman2009inertia,wildeman2016spreading}. The numerical simulations of Wildeman \etal \cite{wildeman2016spreading} are of particular interest for this study, as they show the location of viscous losses within the drops during impacts, and quantify the fraction of energy dissipated as viscous work and internal kinetic energy over the entire spreading step with no-slip and free-slip boundary conditions. According to this model, nearly half of the initial kinetic energy of low-viscosity droplets is lost as viscous dissipation in the lamella during spreading, regardless of the slip length, impact velocity and fluid viscosity.
	
	%At low impact speed, the droplet does not deform much and the kinetic energy is almost integrally converted to surface energy, and then restored back as kinetic energy, which may result in droplets bouncing back \cite{richard2002contact,biance2006elasticity}. 
	The stability of the lamella formed during the impact is also a key concern for DEG. Intuitively, a splashing droplet releases some surface and kinetic energy in the form of ejected daughter droplets. Lamella breakup depends on a competition between a destabilizing suction and lubrication forces that lift it from the substrate, and a restoring capillary force that pulls the liquid back to the bulk of the droplet \cite{xu2005drop}. When the restoring force is overwhelmed by the other two, the lamella detaches and breaks into smaller droplets, resulting in splashing \cite{riboux2014experiments}. Even if the lamella remains stable and recedes, the liquid film itself may have thinned past its own stability limit and rupture into lower-energy liquid islands \cite{dhiman2010rupture}. %When the droplet recoils, the surface energy is converted back to kinetic energy and vibration energy, and a rebound may be observed on super-hydrophobic surfaces when the impact speed is large enough and the droplet speed is not too big \cite{richard2000bouncing,biance2006elasticity}.
	
	In this paper, we investigate the consequences of the hydrodynamic phenomena on the operating conditions and efficiency of DEG devices. An important hypothesis of our work is that the droplet provides no electrical energy during the spreading phase, and that instead the totality of the energy is obtained during the receding of the lamella. This is supported by a thought experiment where one would drop a neutral conducting disc on the DEG. The biased capacitor would release the same amount of energy without any work from the disc. However, one would need to overcome the Coulomb force between the mobile charges induced in the disc and the underlying polymer to detach the disc from the DEG, thereby attesting of an electromechanical conversion during recoil. The paper is organized as follows: starting from the total energy available from a falling droplet, we first evaluate the maximum impact velocity allowed for DEG devices depending on the droplet volume. We then evaluate how this energy is converted into surface energy and turbulent/viscous losses during the spreading phase, followed by the recoil phase. Eventually, we combine our hydrodynamic analysis and the electrical model of Wu \etal \cite{wu2020energy} to quantify the energy efficiency of simulated and experimental DEG devices for various operating conditions, and point out to possible improvements for this technology.

	\section{Methods}
	\subsection{Fluid mechanics}
	\subsubsection{Available energy}
	A falling droplet combines a kinetic energy $\mathcal{K}_0 = \frac{2\pi}{3}a^3\rho {U_0}^2$ and surface energy $\mathcal{V}_0 = 4\pi a^2 \gamma$. Neglecting gravity, the minimum energy of a liquid in contact with a solid surface is obtained when the spherical droplet intersects with the solid surface at the Young contact angle $\cos\theta = \frac{\gamma_{sv}-\gamma_{sl}}{\gamma}$. According to volume conservation, this yields \cite{lubarda2011analysis}:
	\begin{eqnarray}
	&\mathcal{V}_\text{eq} &= \gamma A_\text{eq,cap} - \gamma\cos\theta A_\text{eq,base}, \\
	\text{with }&A_\text{eq,cap} &= 2\pi {R_\text{eq}}^2(1-\cos\theta),\\
	&A_\text{eq,base} &= \pi {R_\text{eq}}^2 \sin^2\theta,\\
	\text{and }& R_\text{eq} &= \frac{2^{2/3}a}{(2-3\cos\theta+\cos^3\theta)^{1/3}},
	\end{eqnarray}
	where $A_\text{eq,cap}$ and $A_\text{eq,base}$ are the cap and base surface area of the droplet at equilibrium, and $R_\text{eq}$ is its radius of curvature. Therefore, the total energy available is obtained by subtracting this lowest possible energy from the initial energy:
	\begin{equation}
	\Emax = \mathcal{K}_0 + \mathcal{V}_0 - \mathcal{V}_\text{eq}. \label{eq: Emax}
	\end{equation}
	\subsubsection{Maximum impact velocity}
	The impact velocity is limited by two factors. On the one hand, the droplet may splash, but even without splashing, the film formed by the impacted droplet may still become unstable and rupture. We first recall the splashing conditions according to Riboux and Gordillo \cite{riboux2014experiments}.	It was experimentally observed that the lamella of impact droplets follow a universal dynamic until an ejection time $t_e$ where splashing and non-splashing droplets will diverge \cite{riboux2014experiments}. Therefore, the droplet state at $t_e$ and particularly the lamella height $H_t$, vertical velocity $V_v$ and recoil speed $V_r$ determine the impact outcome. $\sqrt{t_e}$ is well-approximated by the real positive root of the fourth-order polynomial:
	\begin{equation}
	\frac{\sqrt{3}}{2\Rey} + \frac{\sqrt{{t_e}}}{We} = 1.1 {t_e}^{2}. \label{eq: t_e}
	\end{equation}  
	Obtaining $t_e$ from \eq{eq: t_e} yields the lamella height at the ejection time  $H_t = (a\sqrt{12}/\pi){t_e}^{3/2}$, which then yields the recoil $V_r$ and vertical $V_v$ velocities:
	\begin{eqnarray}
	&V_r &= \sqrt{2\gamma/\rho H_t},\\
	&V_v &= \sqrt{\frac{\ell}{\rho H_t}}\\
	\text{with }&\ell &= K_l\mu_gV_t+K_u\rho_g {V_t}^2H_t,
	\end{eqnarray}
	with $K_l\simeq-\left[6/\tan^2\alpha\right](\log(19.2 \lambda_g/H_t)-\log(1+19.2\lambda_g/H_t))$ and $K_u=0.3$ $\mu_g$ two hydrodynamic coefficients for the suction and lubrication forces that make up the lift force $\ell$, and $\rho_g$, $\mu_g$ and $\lambda_g$ the gas phase density, dynamic viscosity and mean-free path. 
	
	Riboux and Gordillo \cite{riboux2014experiments} postulate that the droplet will splash if the vertical velocity of the forming lamella exceeds the recoil speed by a factor $b_\mathrm{max} = 0.14$, that is:
	\begin{equation}
	V_v \geq b_\mathrm{max} V_r. \label{eq: splash}
	\end{equation}
	
	Even in the absence of splashing, the liquid film formed after impact may still rupture. Surprisingly, even though instability of static liquid films is a well-studied topic \cite{lamb1921statics,taylor1973making,sharma1989dewetting}, film rupture immediately after droplet impact has received much less attention than the more spectacular splashing. Diman and Chandra \cite{dhiman2010rupture} have studied the disintegration of liquid films formed after high-speed collision between a droplet and a wall. While their analysis is rather involved, and depends on the liquid-solid contact angle and the size of defects that trigger the liquid film instability, experimental evidence over a range of contact angles, surface roughness  and liquids suggest that the liquid film will rupture if the droplet impacts a solid wall above some critical Reynolds number $\Rey_c \simeq 5000$:
	\begin{equation}
	\Rey\geq \Rey_c. \label{eq: rupture}
	\end{equation}
	
\begin{figure}
		\includegraphics[width=80 mm]{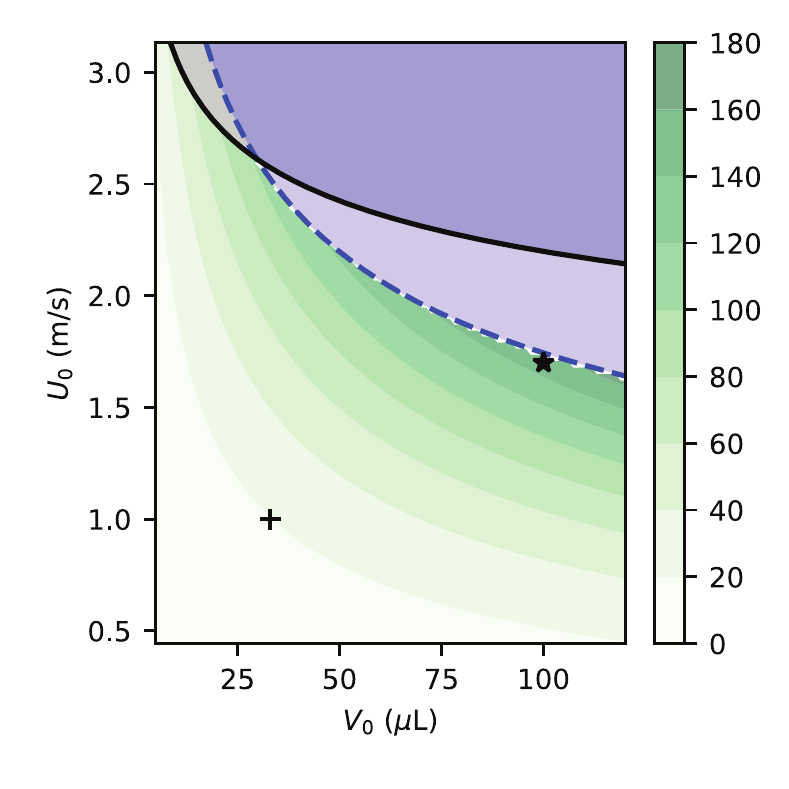}
		\caption{Maximum energy $\Emax$ available from impacting droplets (in $\mu$J) as given by \eq{eq: Emax}. The splashing (solid line) and film-rupturing (dashed line) limits are given by Eqs. \eqref{eq: splash} and \eqref{eq: rupture}. The markers + and $\star$ represent the experiments of Wu \etal \cite{wu2020energy} (energy available 19 $\mu$J) and Xu \etal \cite{xu2020a} (energy available 151 $\mu$J), respectively. \label{fig: max energy}}
	\end{figure}

	The available energy and the limiting velocities for splashing (Eq. \ref{eq: splash}) and film rupturing (Eq. \ref{eq: rupture}) are shown in \fig{fig: max energy}. Note that smaller droplets tend to splash first while larger ones may not splash but their liquid film will rupture nonetheless. The largest theoretical amount of energy while maintaining droplet integrity is obtained for the largest droplets at velocities close to $1.7$ m/s, remarkably close to Xu \etal \cite{xu2020a}.

	\subsubsection{Efficiency during spreading and receding}
	Depending on the impact speed, a sizable fraction of the available energy (Eq. \eqref{eq: Emax}) may be dissipated into turbulent kinetic energy and viscous work during the droplet spreading. The kinetic-to-surface energy conversion efficiency $\eta_s = \frac{\Vmax}{\Emax}$, with $\Vmax = \gamma(1-\cos\theta)\Amax$ the surface energy at maximum spread $\Amax$, can be computed directly from experimental data or numerical simulations, or can be evaluated based on theoretical analyses \cite{ukiwe2005on,pasandideh1996capillary,clanet2004maximal}. This yields the conversion efficiency $\eta_s$:
	\begin{equation}
	\eta_s = \frac{\Vmax}{\Emax}. \label{eq: eta_s}
	\end{equation}

	\begin{figure}
		\includegraphics[width=80 mm]{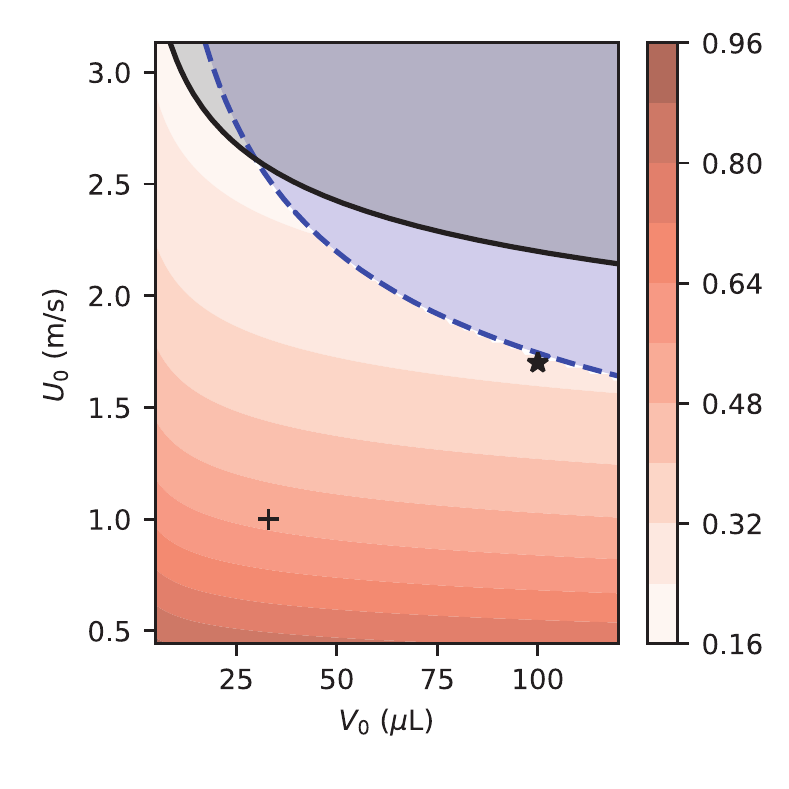}
		\caption{Mechanical to kinetic conversion efficiency $\eta_s$ (\%) according to \eq{eq: a_max}. The splashing (solid line) and film-rupturing (dashed line) limits are given by Eqs. \eqref{eq: splash} and \eqref{eq: rupture}. The markers + and $\star$ represent the experiments of Wu \etal \cite{wu2020energy} (spreading efficiency 17\%) and Xu \etal \cite{xu2020a} (spreading efficiency 30\%), respectively. \label{fig: efficiency}}
	\end{figure}

	$\Amax$ may be determined from from experimental or simulated droplet impacts, or using simplified models (see \cite{ukiwe2005on} and \cite{roisman2009inertia} for critical reviews). Among these models,  Pasandideh-Fard \etal \cite{pasandideh1996capillary} provide a simple and accurate estimate of the maximum spreading diameter of impinging droplets:
	\begin{equation}
	a_\text{max} = a\sqrt{\frac{\We+6}{\frac{3}{2}(1-\cos\theta)+4\We\sqrt{\frac{2}{\Rey}}}}. \label{eq: a_max}
	\end{equation}
	In spite of being derived for highly viscous fluids only, this formula was empirically found to work well for low-viscosity fluids as well \cite{ukiwe2005on,roisman2009inertia}. We refer the reader to Wildeman \etal \cite{wildeman2016spreading} for a more physically-sound model at low viscosity. The resulting energy efficiency obtained by combining Eq. (\ref{eq: eta_s},\ref{eq: a_max}) is shown in \fig{fig: efficiency}.

	During the recoil phase, the surface energy $\Vmax$ is split into 3 terms: the viscous work during recoil $\mathcal{W}_r$, the electrical work during recoil $\mathcal{W}_e$ and the final mechanical energy remaining in the droplet $\Einf$:
	\begin{equation}
	\Vmax =  \mathcal{W}_r + \mathcal{W}_e + \Einf
	\end{equation}
	Experiments with water drops bouncing on super-hydrophobic substrates \cite{richard2000bouncing,biance2006elasticity} suggest that the viscous dissipation is small during the recoil step, even at high impact velocity, and that most of the surface energy is either restored as external kinetic energy or vibration energy. This is confirmed by the following rough estimate of the droplet viscous dissipation. We first assume that the recoil speed scales as $U_r = a_\text{max}/\tau_r$, with $\tau_r$ the contact time between the droplet and the substrate. For super-hydrophobic substrates, the contact time, that is a negligible spreading time plus the receding time, is insensitive to the impact velocity \cite{richard2002contact,kim2001recoiling}. Hence, the recoil time scales proportionally to $\tau_r = \pi\tau_h/\sqrt{2}$ with $\tau_h = \sqrt{\frac{\rho a^3}{\gamma}}$ the oscillation period of a levitated drop \cite{strutt1879vi,kim2001recoiling}. Using $U_r$ as the characteristic recoil velocity, we then compute the volumic shear rate similarly to Pasandideh-Fard \cite{pasandideh1996capillary} and obtain a coarse estimate of the viscous work during recoil:
	\begin{eqnarray}
	&\mathcal{W}_r &\simeq \tau_r  \mu \left(\frac{U_r}{\delta}\right)^2V_0, \label{eq: Wr}\\
	\textbf{with:} & \delta & = \min(h_\text{bl},h_\text{film}),\\
	& h_\text{bl} & = \sqrt{\frac{2\mu\tau_r}{\rho}},\\
	& h_\text{film} & = \frac{V_0}{\Amax}.
	\end{eqnarray} 
	where $h_\text{bl}$ and $h_\text{film}$ are the hydrodynamic boundary layer and  film thickness during recoil. 
	
	\subsection{Electrical model}
	
	\subsubsection{Wu's model}
	By analogy with a biased capacitor, Wu \etal \cite{wu2020energy} have derived the following equation for the charge $q$ driven through the load by the droplet motion:
	\begin{equation}
	\frac{d q}{d t} = \frac{1}{R c_p}\left(\sigma - \frac{q}{A}\right), \label{eq: qdot}
	\end{equation} 
	with $\sigma$ the surface charge of the polymer, $c_p$ the areal capacitance of the polymer and $R = R_L+R_D$ the total resistance of the circuit, including the droplet resistance $R_D$ and the load $R_L$. In Wu's model, $A(t)$ stands for the evolving area of the droplet, but the overlap area of charged polymer in contact with the droplet should be used instead when the polymer charge is non-uniform \cite{wu2020charge}.
	
	The energy output reads $\mathcal{W}_L = \int_{0}^{t_c} R_L \left(\frac{d q}{d t}\right)^2 dt$, where the origin of time $t=0$ is chosen at droplet contact, and $t_c$ is the time when the droplet detaches from the electrode. Using integration by part and substituting \eq{eq: qdot} in this integral yields:
	\begin{eqnarray}
	& \mathcal{W}_L &=  \eta_L \left(\Estat- \Edyn \right), \label{eq: WL}\\
	\text{with:} & \eta_L &= \frac{R_L}{R}, \label{eq: etaL}\\
	& \Estat & = \frac{\sigma q(t_c)}{c_p}, \label{eq: Estat}\\
	& \Edyn & = \frac{1}{2 c_p}\int_{0}^{t_c}\frac{\frac{d q^2}{d t}}{A}dt \label{eq: Edyn}
	\end{eqnarray} 
	Note that the right-hand side of \eq{eq: Edyn} cannot be simplified immediately because there is no one-to-one correspondence between $A(t)$ and $q^2(t)$. \eq{eq: WL} is made of three parts: an efficiency factor $\eta_L$ that accounts for the resistive losses in the liquid, an electrostatic energy $\Estat$ and a contribution that depends on the dynamics of the charges and droplet geometry $\Edyn$. We note that $\Edyn$ is positive only if $q^2$ decreases, meaning that this term acts as a generator only when charges are moving out of the droplet.

	\subsubsection{Available energy}
	The droplet exchanges electrical energy with the load twice. First during the discharge of the bottom electrode in the liquid, and then when the recoil step. As stated in the introduction, we believe that the discharge process goes without energy exchange between the droplet and the DEG: the liquid merely acts as a conductor to release the stored electrostatic energy. 
	Although the discharge step requires no work from the droplet, the droplet-polymer interface becomes increasingly charged thereafter. During the recoil step, the liquid interface area shrinks such that the liquid-polymer interfacial capacitance decreases, which prompts charges to flow back to the bottom electrode. In order to reduce the contact area between oppositely charged surfaces, some electrowetting work must be provided to overcome the electrostatic energy \cite{mugele2005electrowetting,krupenkin2011reverse}. When charges flow back, they provide resistive electrical work through the load but also through the liquid, thereafter called resistive losses:
	\begin{equation}
	\mathcal{W}_\text{R} = (1-\eta_L)\left(\Estat- \Edyn \right). \label{eq: W_R}
	\end{equation} 
	
	Upon sufficient recoil, the droplet eventually  detaches from the top electrode and the current stops flow through the load. Integrating \eq{eq: qdot} shows that the static loss cannot be entirely eliminated unless the droplet area vanishes at $t_c$ (see supplementary information). This can be achieved by using non-uniform surface charges \cite{wu2020charge}. While there is no constraint on the remaining amount of charges (missing charges can be provided from the electrical ground), a highly charged droplet will need to expend more energy to leave the substrate than a neutral one. Therefore, the least charge remain in the droplet, the most energy is available for the load. We will refer to the additional electrostatic expense at the static loss $\mathcal{W}_\text{stat} = \frac{q^2(t_\infty)}{2c_p A(t_\infty)}.$ where $t_\infty$ indicates the time the droplet breaks away from the DEG substrate. By conservation of charge, $q(t_c)=q(t_\infty)$, and by conservation of energy the work provided to change the droplet surface area must compensate exactly the variation of $\mathcal{W}_\text{stat}$, therefore:
	\begin{equation}
	\mathcal{W}_\text{stat} = \frac{q^2(t_c)}{2c_p A(t_c)}. \label{eq: W_stat}
	\end{equation} 
	
	\section{Results and discussion}
	
\begin{figure}
		\includegraphics[width=80 mm]{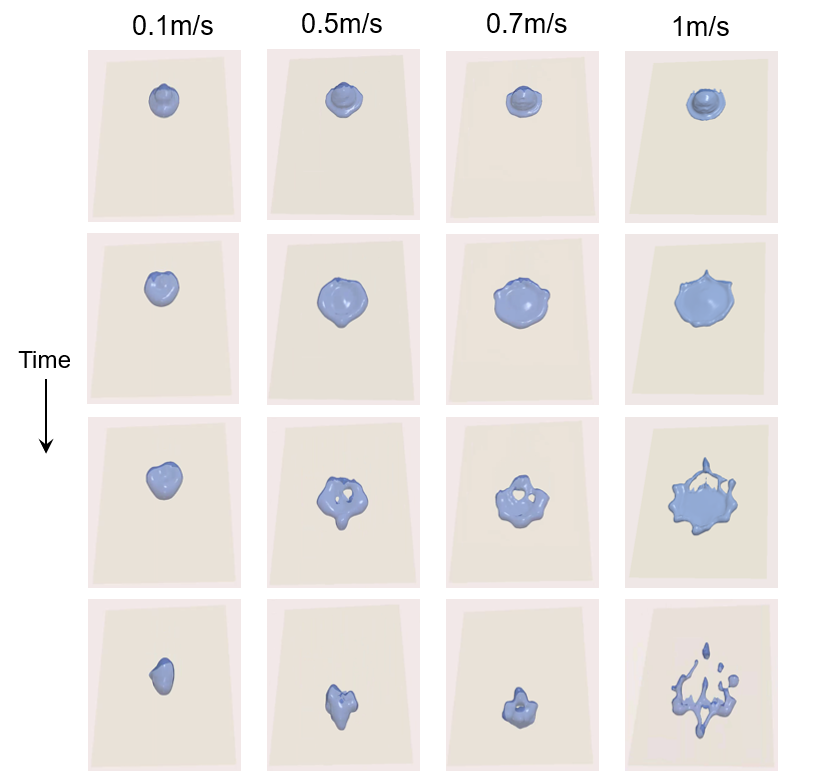}
		\caption{Simulations of droplet impact on Teflon. The liquid film becomes unstable for impact velocities above 1 m/s. See the supplementary information for the simulation parameters and time in pictures. \label{fig: openfoam impacts}}
	\end{figure}

	\fig{fig: max energy} suggests that large droplets yield the maximum energy within hydrodynamic stability limits. Accordingly, we have used OpenFOAM to simulate the impact of $100$ $\mu$l droplets on Teflon with an angle of 45$^o$ at various velocities (\fig{fig: openfoam impacts}). In agreement with the stability bounds, we observe the onset of film rupture (and not splashing) as the impact velocity exceeds 1 m/s ($\Rey \simeq 2000$).

\begin{figure}
		\includegraphics[width=80 mm]{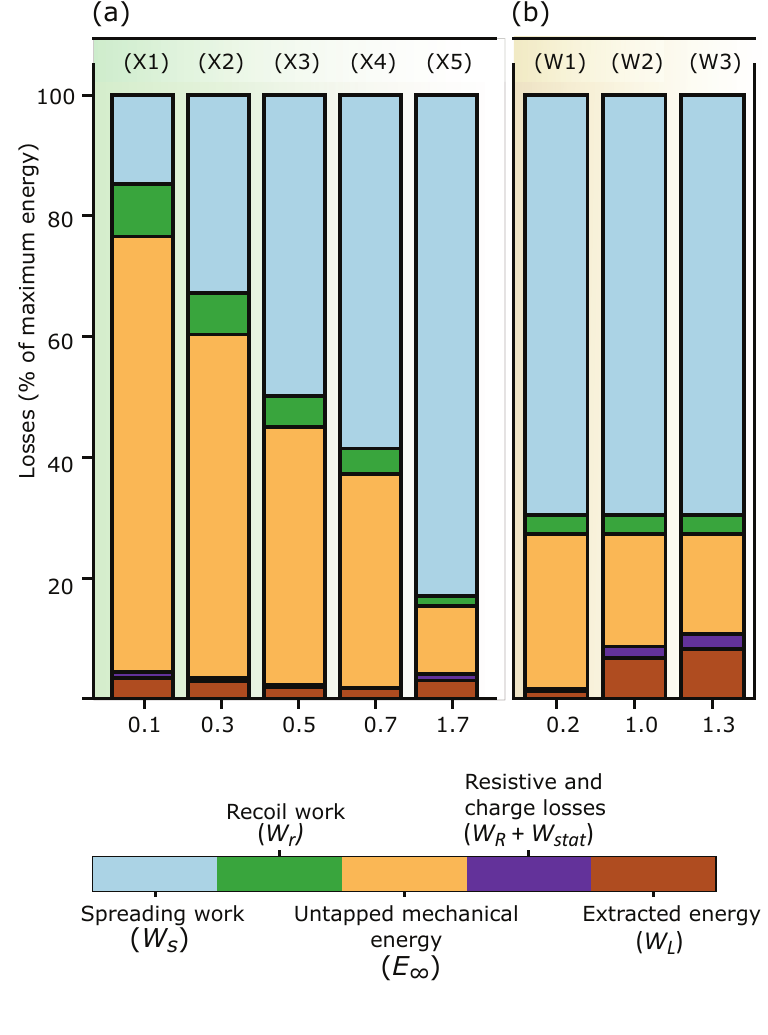}
		\caption{Energy losses at successive stages of the electricity generation as a percentage of the maximum energy $\Emax$ \textbf{(a)}  for 100 $\mu$L droplets with impinging speed ranging from 0.1 to 1.7 m/s, \textbf{(b)} for 33 $\mu$L droplets with varying initial electrical energy $E_0 = \Amax\sigma^2/(2c_p)$.  The symbols (X1..5) and (W1..3) indicate the experimental parameters according to table \ref{tab: parameters}. \label{fig: efficiency histo}}
	\end{figure}

	Based on DEG impact videos \cite{xu2020a,wu2020energy} and our simulations, we evaluate the surface energy $\Vmax = \gamma(1-\cos\theta)\Amax$ available as the impinged droplet spreads to its maximum diameter. The viscous work during recoil is then deduced from \eq{eq: Wr}. Similarly to previous studies \cite{xu2020a,wu2020energy}, we optimize the value of $R_L$ and the delay between impingement and electrical contact to maximize the energy output at given $A(t)$ (obtained from experimental videos or simulations). With these optimized parameters, we compute the energy generated per droplet, together with the static and resistive losses (Eq. \ref{eq: WL}, \ref{eq: W_stat} and \ref{eq: W_R}), which yields the total electrical energy available $\mathcal{W}_e$. The relative shares of each energy contribution are shown in \fig{fig: efficiency histo}.
	
	\begin{table*}
		\caption{\label{tab: parameters} Parameters used for droplet electric generation loss analysis in Figs. \ref{fig: efficiency histo}.}
		\begin{ruledtabular}
			\begin{tabular}{lccccccc}
				Case & Volume & Impact speed & Dielectric & Surface charge & A(t) & $E_0$ & $\Emax$\\
				& ($\mu$l) &(m/s) & & (mC/m$^2$) & & ($\mu$J)&($\mu$J)\\
				
				(X1) & 100 & 0.1 & PTFE (16 $\mu$m) & 0.17 & Simulated & 0.2 & 7.3 \\
				(X2) & 100 & 0.3 & PTFE (16 $\mu$m) & 0.17 & Simulated & 0.2 & 11.3\\
				(X3) & 100 & 0.5 & PTFE (16 $\mu$m) & 0.17 & Simulated & 0.3 & 19.2\\
				(X4) & 100 & 0.7 & PTFE (16 um) & 0.17 & Simulated & 0.4 & 31.2 \\
				(X5) & 100 & 1.7 & PTFE (16 $\mu$m) & 0.17 & Experimental \cite{xu2020a} & 3.4 & 150.1\\
				(W1) & 33 & 1.0 & 
				\begin{tabular}{@{}c@{}}SiO$_2$ (300 nm) and\\ PTFE (900 nm) \end{tabular}
				& 0.35 & Experimental \cite{wu2020energy} & 0.2 & 19.4\\
				(W2) & 33 & 1.0 &
				\begin{tabular}{@{}c@{}}SiO$_2$ (400 nm) and\\ Cytop (400 nm)  \end{tabular}
				& 1.15 & Experimental \cite{wu2020charge} & 1.0 & 19.4\\
				(W3) & 33 & 1.0 & 
				\begin{tabular}{@{}c@{}}SiO$_2$ (400 nm) and\\ Cytop (120 nm)  \end{tabular}
				& 1.80 & Experimental \cite{wu2020charge} & 1.3 & 19.4
				
			\end{tabular}
		\end{ruledtabular}
	\end{table*}
	
	According to \fig{fig: efficiency histo}(a), the conversion efficiency of impinging droplets depends on the droplet size and impact speed. For large droplets impacting at high speed (X5), most of the energy (83\%) is lost as viscous work, in good agreement with \fig{fig: efficiency} and previous studies \cite{wildeman2016spreading}. The recoil consumes less than 2\% of the total energy $\Emax$ by itself. The second largest loss contribution ($\simeq$ 10\% of the total, 74\% of the energy available after recoil) is the difference between the energy after recoil and the electrical energy. This large mismatch is shared across all sizes of droplets regardless of the impact speed, and represents the mechanical energy $\Einf$ remaining after the droplet detaches from the electrode. By analogy with bouncing droplets \cite{richard2000bouncing,biance2006elasticity}, it is likely that this energy is a combination of internal kinetic energy and surface vibration energy. Next, an optimized load allows extracting almost $75\%$ the electrical energy (resistive losses are negligible and the static losses represent approximately 1\% of $\Emax$).
	
	Having identified that most of the energy is dissipated by viscous work and inefficient energy conversion during recoil, we now point to some ways to reduce these losses. The ratio of viscous to capillary work scales as $\Ca = \frac{\mu U_0}{\gamma}$ and looks insensitive to the main dimensions of the surface, which suggests that micropatterns which were successful at reducing the droplet spreading and contact time \cite{liu2014pancake} may not help minimizing the viscous dissipation. However, our simulations suggest that decreasing the impact velocity from $1.7$ m/s (X5) to $U_0=0.1$ m/s (X5) can cut the viscous dissipation during spreading to only 15\% of $\Emax$. This is in line with the improvement of restoring coefficient of bouncing Leidenfrost-levitated droplets at lower speed \cite{biance2006elasticity}. Therefore, while high-speed impacts generate more energy, low-speed impacts are considerably more efficient. Nonetheless, low-speed impacts fail to extract the remaining mechanical energy from the droplet, so that only 5.1\% of $\Vmax$ is converted to electricity when $U_0$ = 0.1 m/s (X1) whereas up to 24\% of $\Vmax$ becomes electricity when $U_0$ = 1.7 m/s (X5).
	
	A tentative interpretation is based on the following rough estimate of the energy conversion \cite{wu2020energy}:
	\begin{eqnarray}
	&\mathcal{W}_L &\simeq 2\left(\frac{\Delta A}{\Amax}\right)^2 E_0, \label{eq: electromech_conv} \\
	\text{with:} & E_0 &= \Amax\sigma^2/(2c_p),
	\end{eqnarray}
	with $E_0$ the electrostatic energy stored in the polymer before the droplet impact and $\Delta A$ the difference of liquid-solid surface area between the time when the droplet connects to the electrode and when it breaks away from it. The factor $2\left(\frac{\Delta A}{\Amax}\right)^2$ represents the charges flowing twice through the load, first during the spontaneous discharge when the droplet contacts the electrode, and then during the reversed electrowetting at recoil. At high impact speed, the liquid motion is irreversible which results in a large $\Delta A$, whereas at low impact speed, the flow motion is essentially reversible so that $\Delta A$ becomes very small which ruins the device overall efficiency.
	
	Similarly, according to \eq{eq: electromech_conv}, reducing the surface capacitance or increasing the surface charge \cite{wu2020charge} increases the ratio of electrical to capillary energy from 5.5\% (W1) to 35\% (W3) as shown in \fig{fig: efficiency histo}(b).
	
	\section{Conclusion}
	
	In this paper, we have used a combination of analytical equations, numerical model and experimental data to map out the energy losses during DEG generation. The energy conversion efficiency of these devices is mainly limited by viscous dissipation and poor capillary to electrical energy conversion. Experimental data and our numerical simulations show that a small fraction of the initial kinetic energy of impinging droplets is converted into surface energy at maximum spread. The remaining energy is lost as shear work and internal kinetic energy. Even though slower impacts provide less peak power, they dissipate much less energy. For applications where the total energy is critical (as opposed to the peak power), slower impact velocities may allow extracting more energy. This may be achieved by cascading generators with small gap height between them, or even harvesting the energy of crawling droplets \cite{cheng2015high}. Furthermore, increasing the device charge is indeed a key element in the path of improving DEG efficiency. By reducing the impact speed and increasing the surface charge, it makes little doubt that DEG efficiency can be improved beyond the current 10\%.
	
	\begin{acknowledgments}
		
		This work was supported by the National Natural Science Foundation of China with Grant No. 51950410582, 61874033 and 61674043, the Science Foundation of Shanghai Municipal Government with Grant No. 18ZR1402600, the State Key Lab of ASIC and System, Fudan University with Grant No. 2018MS003 and 2020KF006. 
		
	\end{acknowledgments}

	\bibliography{dropletEnergy}% Produces the bibliography via BibTeX.
	
\end{document}